\documentclass[twocolumn,aps,superscriptaddress,amsmath,secnumarabic,pdflatex,
amssymb,prl]{revtex4}
\usepackage{graphicx}
\usepackage{bm}
\usepackage{amsmath,amsfonts,amssymb}
\usepackage{xcolor}
\usepackage{color}
\usepackage{wasysym}

\usepackage{graphicx}
\usepackage{multirow}

\begin{document}
\title{ Laser Driving Highly Collimated $\gamma$-ray Pulses for the Generation of $\mu^-\mu^+$ and $e^-e^+$ Pairs in $\gamma-\gamma$ Collider}

\author{J. Q. Yu}
\affiliation{State Key Laboratory of Nuclear Physics and Technology, and Key Laboratory of HEDP of the Ministry of Education, CAPT, Peking University, Beijing, 100871, China}

\author{R. H. Hu}
\affiliation{State Key Laboratory of Nuclear Physics and Technology, and Key Laboratory of HEDP of the Ministry of Education, CAPT, Peking University, Beijing, 100871, China}

\author{Z. Gong}
\affiliation{State Key Laboratory of Nuclear Physics and Technology, and Key Laboratory of HEDP of the Ministry of Education, CAPT, Peking University, Beijing, 100871, China}

\author{A. Ting}
\affiliation{Institute for Research in Electronics and Applied Physics, University of Maryland, College Park, Maryland 20742 USA}

\author{Z. Najmudin}
\affiliation{ The John Adams Institute for Accelerator Science, The Blackett Laboratory, Imperial College London, SW7 2BZ, UK. }

\author{D. Wu}
\affiliation{State Key Laboratory of High Field Laser Physics, Shanghai Institute of Optics and Fine Mechanics, Shanghai, 201800, China}

\author{H. Y. Lu}
\email{hylu@pku.edu.cn}
\affiliation{State Key Laboratory of Nuclear Physics and Technology, and Key Laboratory of HEDP of the Ministry of Education, CAPT, Peking University, Beijing, 100871, China}
\affiliation{Collaborative Innovation Center of Extreme Optics, Shanxi University, Taiyuan, Shanxi, 030006, China.}

\author{W. J. Ma}
\email{wenjun.ma@pku.edu.cn}
\affiliation{State Key Laboratory of Nuclear Physics and Technology, and Key Laboratory of HEDP of the Ministry of Education, CAPT, Peking University, Beijing, 100871, China}

\author{X. Q.~Yan}
\email{x.yan@pku.edu.cn}
\affiliation{State Key Laboratory of Nuclear Physics and Technology, and Key Laboratory of HEDP of the Ministry of Education, CAPT, Peking University, Beijing, 100871, China}
\affiliation{Collaborative Innovation Center of Extreme Optics, Shanxi University, Taiyuan, Shanxi, 030006, China.}
\affiliation{Shenzhen Research Institute of Peking University, Shenzhen 518055, China.}


\begin{abstract}

A scheme to generate highly collimated $\gamma$-ray pulse is proposed for the production of muon and electron pairs in $\gamma-\gamma$ collider. The $\gamma$-ray pulse, with high conversion efficiency, can be produced as the result of electron phase-locked acceleration in longitudinal electric field through the interaction between an ultra-intense laser pulse and a narrow tube target. Numerical simulation shows that 18\% energy of a 10-PW laser pulse is transferred into the forward $\gamma$-rays in a divergence angle less than $ 3^\circ$. The $\gamma$-ray pulse is applied in $\gamma-\gamma$ collider, in which muon pairs can be produced and electron pairs can be enhanced by more than 3 orders of magnitude. This scheme, which could be realized with the coming 10PW class laser pulses, would allow the observation of a $\gamma-\gamma$ collider for electron and muon pairs in laboratory.

\end{abstract}

\maketitle

The laser intensity above $10^{23}$W/cm$^2$, which will be accessible by the under construction laser facilities \cite{ELI,XCELS}, at which a large portion of the laser pulse energy can be transferred into $\gamma$-rays \cite{Capdessus2013,Nakamura2012,Brady2012} in the laser matter interaction at the intensity. The $\gamma$-ray source with high conversion efficiency has attracted particular interest due to its wide applications \cite{Ridgers2012,Burke1997,Pike2014,Ribeyre2016,Zhu2016,Li2015,Bulanov2015,Meszaros2013,ELI}, among which the research on quantum electrodynamics (QED) should be one of the most important. In the interaction, although the laser field strength is several orders of magnitude lower than the Schwinger field $E_{crit} = 1.3 \times 10^{18} $ Vm$^{-1}$ \cite{Schwinger1951}, QED effects \cite{Bell2008,Piazza2012,Ridgers2012,Gong2016} become significant.

According to QED, matters can be created from light through Breit Wheeler (BW) \cite{Breit1934,Gould1967,Telnov1995,Burke1997,Ridgers2012} process. In the linear BW process, only two photons involved in one collision, $e^-e^+$ pairs could be generated through $\gamma$-photon \cite{Pike2014} or $\gamma-\gamma$ \cite{Ribeyre2016}, and the next challenges should be the generations of much heavier leptons such as muon and tau pairs. The 10-PW laser pulse can produce a huge number of $\gamma$-rays \cite{Brady2012,Brady2013,Brady2014,Ridgers2012} whose energy are much higher than the rest mass of muon. It is possible to generate muon pairs in the $\gamma-\gamma$ collider driving by 10-PW laser pulses. However, $\mu^-\mu^+$ pair generation in $\gamma-\gamma$ collider has never been proposed due to the large divergence angle ($\sim 30^\circ$) of the $\gamma$-rays \cite{Brady2012,Brady2013,Brady2014,Ridgers2012}.  Although efforts had been made to enhance the collimation of $\gamma$-rays \cite{Stark2016,Huang2017}, there is no distinguished enhancement on the collimation.

In this Letter, highly collimated $\gamma$-ray pulse is generated in the interaction between a 10-PW laser pulse and a narrow tube target. The electron collimation can be improved by phase-locked acceleration of longitudinal electric field $E_x$. The $\gamma$-rays are generated near the tube inner boundary where the electrons are wiggled by the space charge field $E_{ys}$ and self-generated field $B_{zs}$.  Particle-in-Cell (PIC) simulation shows that the forward $\gamma$-rays which occupy 18\% of the laser pulse energy are collimated into a divergence angle less than $3^\circ$. The brilliance could be two orders of magnitude higher than the previous studies \cite{Brady2012,Brady2013,Brady2014,Stark2016,Huang2017,Ridgers2012,Nakamura2012}. A $\gamma-\gamma$ collider, which could produce muon pairs, is proposed using the $\gamma$-ray pulses, in which the production of electron pair can be enhanced by 3 orders of magnitude than before \cite{Ribeyre2016,Pike2014}. Hence, the $\gamma$-ray pulses will allow the observation of the creation of $\mu^-\mu^+$ and $e^-e^+$ pairs from pure light.

In the narrow tube target, the electron dynamics, which impacts the $\gamma$-ray generation, is affected by the traveling longitudinal electric field \cite{Robinson2013} generated in the confined space \cite{Xiao2016}. However, the underlying physics of electron dynamics and radiation in the narrow tube target is never clearly described although it has been intense applied \cite{Yu2012,Jiang2016,Xiao2016,Ji2016,Yi2016}. For simplicity, the electron dynamics in a plane wave $E_y = E_0\sin\phi$, $B_z = B_0\sin\phi$ with longitudinal field $E_x = \rho E_0 \cos\phi$ is considered, where $\phi=t-x$. The electron motion equations can be written as $dp_y/dt = -E_y + v_xB_z$, $dp_x/dt = -E_x - v_yB_z$ and $d\gamma / dt = -v_yE_y - v_xE_x$, where $v_y = dy / dt$, $v_x = dx / dt $ and $\gamma=\sqrt{1+p_x^2+p_y^2}$. The vector potential is $a_y = a_0 \cos(\phi)$ and $E_y = - \partial a_y / \partial t$, $B_z = \partial a_y / \partial x$. Here $E$, $B$, $t$, $x$, $p$ and $a$ are normalized by $m_e\omega_lc/e$, $m_e\omega_l/e$, ${\omega_l}$, $\omega_l/c$, $m_ec$ and $m_ec/e$, $\omega_l$ is laser frequency, $c$ is the light speed in vacuum and $m_e$ is the electron mass at rest. The electron, initially at rest, is pulled into the laser fields at initial phase $\phi_0$. Then, $d(p_x - \gamma)/d\phi = -E_x$,
\begin{equation}
 p_x = \frac{1+a^2_0 (\cos \phi - \cos \phi_0 )^2 - [1+\rho a_0(\sin \phi - \sin \phi_0 )]^2}{2[ 1 + \rho a_0 (\sin \phi - \sin \phi_0)]}.
    \label{px}
\end{equation}
\begin{equation}
 p_y = a_y(\phi) - a_y(\phi_0) = a_0 (\cos\phi - \cos\phi_0).
    \label{py}
\end{equation}

 \begin{figure}
 \centering
 \includegraphics[width= 0.45\textwidth]{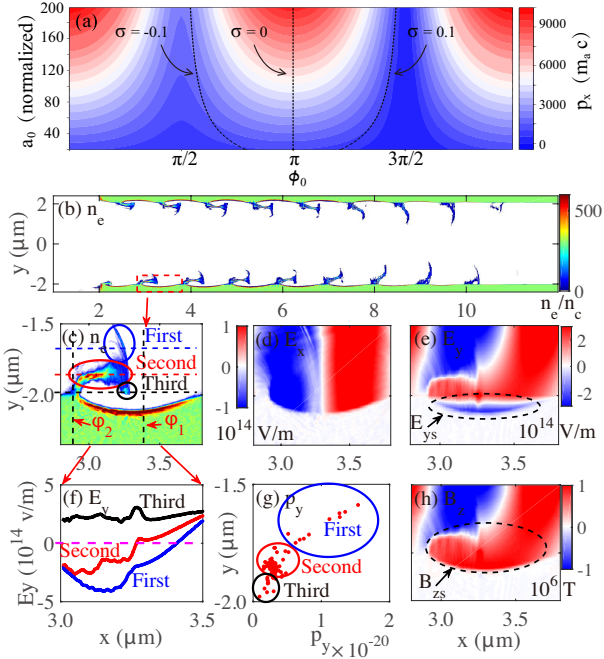}
 \caption{(a) Electron momenta at $t=40\pi$ for different laser amplitude $a_0$ and electron initial phases $\phi_0$. The motion equations are numerically solved using Boris solver with $\Delta t=\pi/10000$. The black lines are theoretic results of equation (\ref{phi}). Simulation results at 40 fs, (b) electron density $n_e$ and (c) the electrons which are pulled into the narrow space can be classified into three groups, (d) the longitudinal electric field $E_x$, (e) the transverse electric field $E_y$ and (h) the magnetic field $B_z$ near the inner boundary of the tube, (f) the $E_y$ whose transverse positions are corresponded to the dotted lines of same color in (c) expressed on the electrons in different groups, (g) the phase-space ($y-p_y$) of the electrons shown in (c). }
 \label{field}
\end{figure}

Equation (\ref{px}) implies that $p_x \to \infty$ and $d\phi/dt \to 0$, when $ 1 + \rho a_0 (\sin \phi - \sin \phi_0) \to 0 $. To have small transverse momentum, $\Delta\phi=\phi-\phi_0 \to 0$ should be satisfied according to Eq. (\ref{py}). Thus,
\begin{equation}
1 + \rho a_0 [\sin (\phi_0 + \Delta \phi) - \sin \phi_0] \simeq 1 + \rho a_0 \Delta \phi \cos \phi_0 \simeq 0,
\label{con1}
\end{equation}

\begin{equation}
\frac{p_y}{a_0}=\cos(\phi_0 + \Delta \phi) - \cos \phi_0\simeq-\Delta \phi\sin\phi_0=\sigma.
\label{con2}
\end{equation}
From Eq. (\ref{con1}) and (\ref{con2}), the locking phase reads
\begin{equation}
\label{phi}
\phi_0 = (2n + 1)\pi + \arctan(\sigma\rho a_0),
\end{equation}
where $\sigma$ is a small quantity and $n=0,\pm 1,\pm 2,\cdots$. Figure \ref{field}(a) gives the electron longitudinal momenta at $t = 40\pi$ for different $a_0$ and $\phi_0$ by assuming $\rho =0.39$. Electrons with $\phi_0 \sim (2n+1)\pi$ are accelerated to high energy but small transverse momenta as $\sigma\sim0$. The maximum longitudinal momenta are proportional to $a_0$. The lines with $\sigma=\pm0.1$ are approximative the boundaries of the phase-locked acceleration phases. For $a_0 > 200$ here, most of the electrons can experience phase-locked acceleration whose phase ranges from $\phi_1 \approx (2n+1/2)\pi$ to $ \phi_2 \approx (2n+3/2)\pi$ as shown in Fig. \ref{field}(a).

In laser wakefield acceleration (LWFA) \cite{Esary2009}, electrons are collimated to small divergence. However, the energy efficiency is much smaller than phase-locked acceleration \cite{Xiao2016} due to acceleration gradient. In the direct laser acceleration (DLA) \cite{Pukhov1999,Arefiev2012,Robinson2013,Gibbon2005}, high energy conversion efficiency can be achieved. While, the electron divergence is typically large. Hence, comparing with LWFA and DLA \cite{Esary2009,Pukhov1999,Arefiev2012,Robinson2013,Gibbon2005}, phase-locked acceleration takes the advantages of higher efficiency and better collimation.

To model the generation of $\gamma$-rays, two-dimensional PIC simulations are performed with EPOCH \cite{Arber2015,Ridgers2014,Duclous2010}. A 10-PW p-polarized laser pulse with Gaussian spatial and sin$^2$ temporal profiles, duration of 30 fs and focal spot diameter of 2.0 $\mu m$ at FWHM, irradiates at a gold narrow tube from the left side. The laser intensity $3.2\times10^{23} $W/cm$^2$, corresponding to $a_0 \approx 480$, would be available in ELI \cite{ELI}. The tube $e^-$ / Au$^{+69}$ density is set to $276n_c$ / $4n_c$, where $n_c $ is the critical density. The tube wall, whose thickness is $400\, nm$ and inner diameter is 4 $\mu m$, is located 2-270 $\mu m$. The simulation box is 271 $\mu$m in longitudinal direction (x) and 10 $\mu$m in transverse direction (y). 5 macro-ions and 100 macro-electrons are initialized in each cell whose size is $dx=dy=5\, nm$.

Electrons are pulled into the narrow space and pushed into the tube wall by $E_{yL}$. High density ($\sim 500n_c$) electron bunches, and $E_{ys}$ and $E_{xs}$ are generated as shown in Fig. \ref{field}(b \& c) and Fig. \ref{field}(d \& e). The transport of the electron bunches results in the generation of $B_{zs}$ \cite{Stark2016} as shown in Fig. \ref{field}(h). The laser fields and the secondary fields ($E_{ys}$, $E_{xs}$ and $B_{zs}$) compose the fields shown in Fig. \ref{field}. Comparing the locations of the electron bunches, $E_y$, $E_x$, $\phi_1$ and $\phi_2$ marked in Fig. \ref{field}(c), results demonstrate that almost all the electrons dragged into the narrow space can be locked into the acceleration phase of $E_x$, which is consistent with the theory.

In the narrow space, the dragged electrons can be classified into three groups as shown in Fig. \ref{field}(c \& g). The transverse force $F_\perp$ exerted on the electrons is
\begin{equation}
 F_{\perp v} \approx  -e E_y + v_x B_z,
 \label{Fyv}
\end{equation}
where $\vec E_y  =  \vec E_{ys} + \vec E_{yL}$ and $\vec B_z  =  \vec B_{zs} + \vec B_{zL}$.

In the first group, $E_{yL} \gg E_{ys}$ and $B_{zL} \gg B_{zs}$, the electron dynamics can be described by our theory as the laser fields dominate the movement. Due to initial large $v_y$ as shown in Fig. \ref{field}(g), the electrons will cross the laser axis and move toward the other side of the tube where most of them could be trapped again by $E_{xL}$. Hence, the electrons in the first group could be trapped and accelerated twice. In the second group, $B_{zs}$ could cancel the local space charge field, $E_{ys} \approx cB_{zs}$. $E_y \approx cB_z$ can be still satisfied as shown in \ref{field}(e \& h). As $F_{\perp v} \to 0$, the electrons can be trapped by $E_x$ for a long time which follows our theory as well. After crossing the phase of $E_{y}$, the electrons will be pulled back to the tube wall by $F_{\perp v}$. In the third group, the electrons which are exposed to larger $E_{ys}$ than $E_{yL}$ as shown in Fig. \ref{field}(f) will be easily pulled back to the tube wall by $E_{ys}$. Thus, the electrons in this group can only contribute to low energy radiation due to shorter acceleration time. The above discussions show that the electrons in the first two groups can be phase-locked accelerated by $E_x$ for a long time.

\begin{figure}
 \centering
 \includegraphics[width=0.45\textwidth]{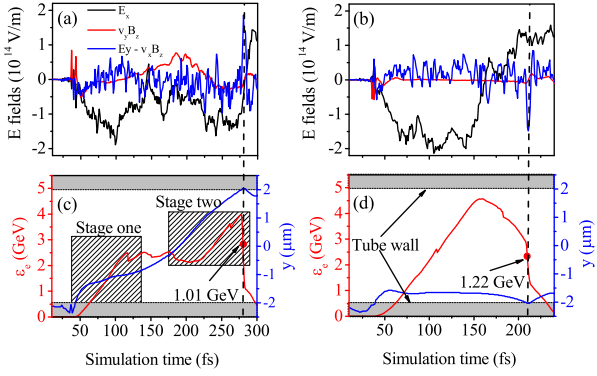}
 \caption{  (a) / (b) The fields exert on the electron, (c) / (d) the kinetic energy history and the track of the corresponding emitting electron in the first group / the second group.}
 \label{detail}
\end{figure}

\begin{figure}
 \centering
 \includegraphics[width=0.45\textwidth]{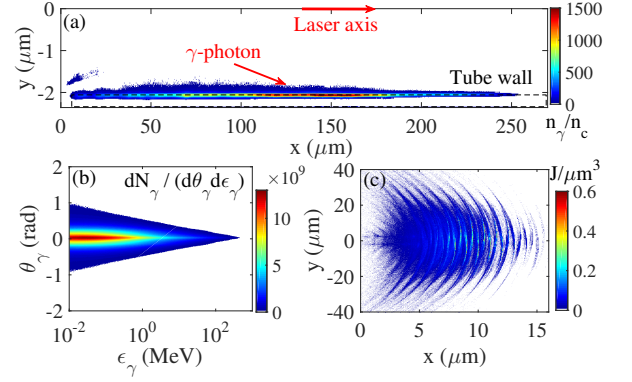}
 \caption{ PIC simulation results at 900 fs, showing the photons with kinetic energy $ > 100$ keV only. (a) The emitting positions of the $\gamma$-rays, only half tube is plotted to show it clear, (b) the spectral and angular distribution of the $\gamma$-ray pulse, and (c) the energy density distributions of the $\gamma$-ray pulse.}
 \label{result}
\end{figure}

Figure \ref{detail}(a \& b) show the fields exerting on the test electrons in the first two groups, and the electron which can be accelerated to several GeV in a distance $< 100 \mu m$ as shown in Fig. \ref{detail}(c \& d) is mostly contributed from $E_x$ instead of $E_y$ and $B_z$. The energy of the emitted $\gamma$-ray photon \cite{Duclous2010,Ridgers2014} could be estimated as $h\nu \approx 0.44\eta\gamma m_ec^2$, where $\eta \approx \gamma \vert F_\perp \vert /(eE_{crit})$. For the generation of $\gamma$-rays, the emitting electrons should be accelerated to high energy before being wiggled. In the tube wall, the direction of $E_y$ turns reversal as circled in Fig. \ref{field}(e), the transverse force $F_\perp$ becomes
\begin{equation}
 F_{\perp w} \approx  -e E_{ys} (1 + \frac{v_x}{c} )  \approx -2 e E_{ys}.
 \label{Fyw}
\end{equation}
Hence, $F_{\perp w}$ is much larger than $F_{\perp v}$. The electrons will be pulled back to the tube wall, where the larger $F_{\perp w}$ will significantly enhance the production of $\gamma$-rays \cite{Jackson} as shown in Fig. \ref{detail}(c \& d) and Fig. \ref{result}(a). All the above characters indicate that the electron movement and acceleration, together with the acceleration in two stages marked in Fig. \ref{detail}(c), well follow the above discussion.

At 900 fs in the simulation, $\sim90\%$ of the laser energy is depleted and the energy conversion $\eta_\gamma$ from the laser to the forward $\gamma$-rays which occupy more than 99\% of the $\gamma$-rays is $ \sim18\%$. Most of the $\gamma$-rays are generated near the tube inner boundary as shown in Fig. \ref{result}(a). The photons are highly collimated to $\theta_\gamma \sim 3^\circ$ as shown in Fig. \ref{result}(b). The $\gamma$-ray pulse duration is about 21 fs (6.3 $\mu m$) and the focal spot diameter is about 25 $\mu m$ at FWHM as shown in Fig. \ref{result}(c). The average photon brilliance, as shown in table \ref{table1}, is much higher than the reported results \cite{Ridgers2012,Brady2013,Stark2016,Huang2017}. It is also found that $\eta_\gamma$, brilliance and the collimation increase with $I$, and 31\% of the laser pulse energy is transferred into $\gamma$-ray pulse collimated into $2^\circ$ at $I = 1.28\times10^{24}$W/cm$^2$ (40 PW laser pulse).

The full 3D condition is simulated by reducing the target length to 50 $\mu m$, in which the results show same characters as 2D. In the experiment, one can use much thicker tube but same diameter as the thickness does not affect the result. In addition, technical methods are needed to minimize the influence of pre-pulse and to inject the laser pulse into the narrow space.

\begin{figure}
 \centering
 \includegraphics[width=0.45\textwidth]{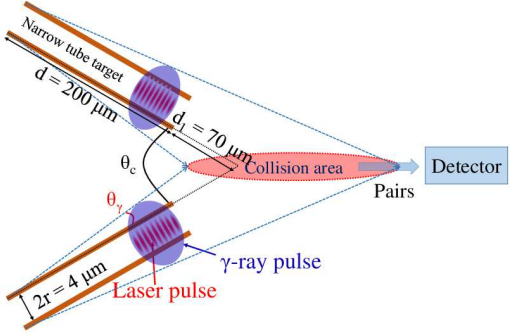}
 \caption{ Schematic of $\gamma-\gamma$ collider for the generations of $e^-e^+$ and $\mu^-\mu^+$ pairs.}
 \label{setup}
\end{figure}

For the highly collimated $\gamma$-ray pulses, one of the most important application is $\gamma-\gamma$ collider. Figure \ref{setup} shows the setup of $\gamma-\gamma$ collider, in which the $\gamma$-ray pulses are generated from laser driving narrow tube targets. The $\gamma$-ray pulses collide in a angle of $\theta_c$ which should fulfill the relation $\theta_c < 180^\circ - 2\theta_\gamma$ to make sure that most of the $\mu^-\mu^+$ and $e^-e^+$ pairs are generated in the forward direction where a detector is placed. $d$, the length of the targets, is fixed to be 200 $\mu m$ as almost all the photons are generated in 70-200 $\mu m$ as shown in Fig. \ref{result}(a). $r = 2 \, \mu m$ is radius of the targets. A distance $d_1$ from target end to the collision point is considered to minimize the e$^-$e$^+$ pairs produced from multi-photon BW process \cite{Ridgers2012}. As the laser intensity decreases significantly after the transporting in vacuum for a distance of $d_1$. According to the Eq (6) in the ref. \cite{Ribeyre2016}, the pair number from BW process can be estimated by

\begin{equation}
 N_{\pm} \sim {\sum_{i=1}^{N_{\gamma_1}}}  \sum_{j=1}^{N_{\gamma_2}} \frac{\sigma_{\gamma_1\gamma_2}}{S_c},
    \label{pair}
\end{equation}
where $N_{\gamma_1}$ and $N_{\gamma_2}$ are the photon number in the $\gamma$-ray pulses. The distance from the source location to the collision point is about $200 \mu m$. The size of collision area $S_c$, which should be doubled for circular-polarized laser pulse, can be expressed as $S_c = \pi(d\sin\theta_\gamma/\sin( 0.5\theta_c + \theta_\gamma ) + (2r\cos\theta_\gamma + d\sin\theta_\gamma)/\sin(0.5\theta_c - \theta_\gamma))^2/8$. The photon-photon cross section $\sigma_{\gamma_1\gamma_2}$  \cite{Gould1967} is
 \begin{equation}
 \sigma_{\gamma_1\gamma_2} = \frac{\pi}{2} r_c^2(1-\beta^2)[(3-\beta^4)\ln\frac{1+\beta}{1-\beta} - 2\beta(2-\beta^2)],
    \label{cross-section}
\end{equation}
where $r_c$ is the classical muon (or electron) radius, $\beta=(1-1/s)^{\frac{1}{2}}$, $s = \varepsilon_{\gamma_1} \varepsilon_{\gamma_2} (1-\cos\theta_c) /2m_l^2c^4 $, $ \varepsilon_{\gamma_1} $ and $ \varepsilon_{\gamma_2}$ are the photon energies, $m_l$ is the rest mass of the leptons. $s >1$ is the threshold condition for pair generation.

To solve Eq (\ref{pair}) and (\ref{cross-section}), macro-photons which are produced from the PIC simulation are used to reduce the data size. All the photons are assumed to collide in $\theta_c$ which is acceptable for highly collimated $\gamma$-ray here. The photon spectra is divided into 500 segments in the step of $\log_{10} (\varepsilon_\gamma (max))/500$, and the photons in the same segment are assumed to collide with the other pulse in the same condition.

For the generation of $\mu^-\mu^+$ pairs, $r_c = e^2/m_\mu c^2$, where $m_\mu$ is muon mass at rest. From Eq (\ref{pair}) and (\ref{cross-section}), it is found that the muon number is about 1.6 per shot from head-on collision in the case of 10 PW laser pulse, while can be enhanced to 160 by using 40 PW laser pulses. For e$^-$e$^+$ pairs production, $r_c$ is changed to $e^2/m_e c^2$, where $e$ is elementary charge. Through the numerical solution of Eq (\ref{pair}) and (\ref{cross-section}), $3.4 \times 10^8$ ($2.8 \times 10^{10}$ ) pairs can be generated from head-on ($\theta_c = 180^\circ$) collision with the $\gamma$-ray pulses generated from 10 PW (40 PW) laser pulses. The pair number is 3 orders of magnitude higher than the reported results \cite{Ribeyre2016,Pike2014}. The effects of $\theta_c$ is explored to see the robustness of this method. It is found that the pair number is significantly affected by $\theta_c$ as shown in table \ref{table1}. To observe strong signal by the detector, optimal value of $\theta_c$ should be larger than $90^\circ$ and less than $180^\circ - 2\theta_\gamma$. Meanwhile, the pairs generated from trident process \cite{Chen2009}, whose optimal thickness is much thicker than the tube wall, can be ignored. Hence, the highly collimated $\gamma$-ray pulses would allow the observation of muon and electron pairs generation in $\gamma-\gamma$ collider in laboratory.

\begin{table}
\caption{ The average photon brilliance $Br$ (photons/s/mm$^2$/mrad$^2$/0.1\%BW) at 270$\mu m$ away from the irradiating point, the photon number ($>100 $keV) $N_\gamma$ in the $\gamma$-ray pulses, the divergence angle $\theta_\gamma$ of the $\gamma$-ray pulses,
the pair numbers of e$^-$e$^+$ and $\mu^-\mu^+$ generated from the $\gamma-\gamma$ collider driven by different laser pulse and collision in different angle $\theta_c$.}
\begin{tabular}{c c c c c}
\hline
\hline
      &  \multicolumn{2}{c}{10 PW}  &    \multicolumn{2}{c}{40 PW}     \\
\hline
          $\theta_\gamma$     &  \multicolumn{2}{c}{$\sim 3^\circ$}      &   \multicolumn{2}{c}{$\sim 2^\circ$}   \\

  $N_\gamma$ ($>$ 100 keV)    &  \multicolumn{2}{c}{$9.3\times 10^{13}$} &   \multicolumn{2}{c}{$6.33\times 10^{14}$}   \\

           $Br$ at 0.5 MeV    &  \multicolumn{2}{c}{$1.5\times 10^{25}$} &  \multicolumn{2}{c}{$1.4\times 10^{26}$}   \\

           $Br$ at 100 MeV    & \multicolumn{2}{c}{$4.8\times 10^{23}$}  &  \multicolumn{2}{c}{$4.5\times 10^{24}$}   \\
\hline
                              &    $\mu^-\mu^+$    &      $e^-e^+$         &    $\mu^-\mu^+$  &        $e^-e^+$       \\
\hline
    $\theta_c = 180^\circ$    &            1.3     &     $3.4\times 10^8$  &         160.5    &  $2.8\times 10^{10}$  \\

    $\theta_c = 150^\circ$     &            1.1     &     $3.0\times 10^8$  &         137.7    &  $2.5\times 10^{10}$  \\

    $\theta_c = 120^\circ$    &            0.64    &     $2.2\times 10^8$  &         85.1     &  $1.8\times 10^{10}$  \\

    $\theta_c = 90^\circ$     &            0.23    & $1.2\times 10^8$      &         34.2     &   $9.4\times 10^9$    \\

    $\theta_c = 60^\circ$     &           0.035     &   $3.9\times 10^7$    &         6.7      &   $3.1\times 10^9$    \\
\hline
\hline
\end{tabular}
 \label{table1}
\end{table}

In conclusion, the electron phase-locked acceleration of longitudinal field is illustrated theoretically to resolve the dynamics of energetic electrons in the interaction between an ultra-intense laser pulse and a narrow tube target. Energetic electrons are accelerated efficiently by longitudinal electric field and radiate strongly near the inner boundary of the tube, resulting in the generation of a highly collimated $\gamma$-ray pulse. The brilliance of the forward $\gamma$-ray pulse, from PIC simulation of 10-PW laser pulse, is two orders of magnitude higher than the reported results due to high conversion efficiency up to 18\% and small divergence angle less than $ 3^\circ$. The $\gamma-\gamma$ collider, in which the $\mu^-\mu^+$ pairs can be created from pure light and production of $e^-e^+$ pairs can be enhanced by 3 orders of magnitude, is proposed by using the $\gamma$-ray pulses. Such highly collimated $\gamma$-ray pulse which could be realized with coming 10PW laser pulses, would benefit the research on quantum electrodynamics, the production of isotopes and laboratory astrophysics.

\begin{acknowledgments}
This work has been supported by the National Basic Research Program of China (Grant Nos. 2013CBA01502, 11475010), NSFC (Grant Nos.11535001, 11575011), National Grand Instrument Project (Grant No. 2012YQ030142), National Key Research and Development Program of China (Grant No. SQ2016zy04003194) and China Postdoctoral Science Foundation (Grant Nos. 2016M600007, 2017T100009). The numerical simulations are carried out in Max Planck Computing and Data Facility. The PIC code Epoch was in part funded by the UK EPSRC grants EP/G054950/1. J. Q. Yu and R. H. Hu contributed equally to this work.
\end{acknowledgments}

\bibliography{aa}

\end{document}